\documentclass[twocolumn,aps,pra,showpacs,floatfix,superscriptaddress]{revtex4-1}
\usepackage{graphicx}
\usepackage{amsmath}
\usepackage{bm}
\usepackage{color}
\usepackage{dcolumn}
\usepackage{hyperref}

\begin{document}


\newcommand{\etal}{{\it et al.}}
\newcommand{\abi}{{\it ab initio}}
\newcommand{\fref}[1]{Fig.~\ref{#1}}
\newcommand{\Fref}[1]{Figure \ref{#1}}
\newcommand{\sref}[1]{Sec.~\ref{#1}}
\newcommand{\Eref}[1]{Eq.~(\ref{#1})}
\newcommand{\tref}[1]{Table~\ref{#1}}
\newcommand{\rtw}{\longrightarrow}
\def\veps{\varepsilon}
\newcommand{\cm}{\ensuremath{\mathrm{cm}^{-1}}}

\newcommand{\cmt}[1]{[\![#1]\!]}
\newcommand{\au}{\ensuremath{\mathrm{a.u.}}}
\newcommand{\pprime}{\ensuremath{\prime\prime}}
\newcommand{\n}{\ensuremath{\left(n+{\textstyle\frac{1}{2}}\right)}}
\newcommand{\np}{\ensuremath{\left(n^\prime+{\textstyle\frac{1}{2}}\right)}}
\newcommand{\sr}[1]{$^{#1}$Sr$_2$}
\newcommand{\yb}[1]{$^{#1}$Yb$_2$}
\newcommand{\cs}[1]{$^{#1}$Cs$_2$}
\newcommand{\g}{\ensuremath{{^1\Sigma_g^+}}}
\newcommand{\gx}{\ensuremath{X\,{^1\Sigma_g^+}}}
\newcommand{\dem}{DeMille \etal}
\newcommand{\zel}{Zelevinsky \etal}
\newcommand{\ger}{Gerber \etal}
\newcommand{\ep}{electron-to-proton mass ratio}

\newcommand{\NZIAS}{
Centre for Theoretical Chemistry and Physics,
New Zealand Institute for Advanced Study,
Massey University, Auckland 0745, New Zealand}

\newcommand{\UNSW}{
School of Physics, University of New South Wales, Sydney 2052, Australia}

\title{Effect of $\alpha$ variation on the vibrational spectrum of \sr{}}

\author{K. Beloy}
\affiliation{\NZIAS}

\author{A. W. Hauser}
\affiliation{\NZIAS}

\author{A. Borschevsky}
\affiliation{\NZIAS}

\author{V. V. Flambaum}
\affiliation{\UNSW}
\affiliation{\NZIAS}

\author{P. Schwerdtfeger}
\affiliation{\NZIAS}

\date{\today}

\begin{abstract}
We consider the effect of $\alpha$ variation on the vibrational spectrum of Sr$_2$ in the context of a planned experiment to test the stability of $\mu\equiv m_e/m_p$ using optically trapped Sr$_2$ molecules [Zelevinsky \etal, Phys.~Rev.~Lett.~{\bf 100}, 043201; Kotochigova \etal, Phys.~Rev.~A {\bf 79}, 012504]. We find the prospective experiment to be 3 to 4 times less sensitive to fractional variation in $\alpha$ as it is to fractional variation in $\mu$.
Depending on the precision ultimately achieved by the experiment, this result may give justification for the neglect of $\alpha$ variation or, alternatively, may call for its explicit consideration in the interpretation of experimental results.
\end{abstract}


\maketitle

\section{Introduction}

In the endeavor to understand nature on its most fundamental level, physicists are striving for a description of all fundamental forces within a single unified theory. Some promising theories suggest that observable quantities such as the \ep{} $\mu\equiv m_e/m_p$ or the fine structure constant $\alpha\equiv e^2/\hbar c$ may not have fixed values \cite{Uza03}. A detected drift in $\mu$ or $\alpha$ could thus provide valuable insight into the fundamental workings of nature beyond our current understanding.
To date, laboratory measurements have verified the stability of $\mu$ and $\alpha$ on the fractional level of $10^{-14}$ \cite{SheButCha08} and $10^{-17}$ \cite{RosHumSch08etal} per annum, respectively.
More stringent laboratory tests are further motivated by recent evidence of a spatial gradient in the value of $\alpha$ based on an analysis of quasar absorption spectra \cite{WebKinMur11etal}. It has been suggested that the Earth's motion relative this gradient may lead to measureable effects in the laboratory \cite{BerFla10}.

\begin{figure}[t]
\begin{center}
\includegraphics*[scale=0.44]{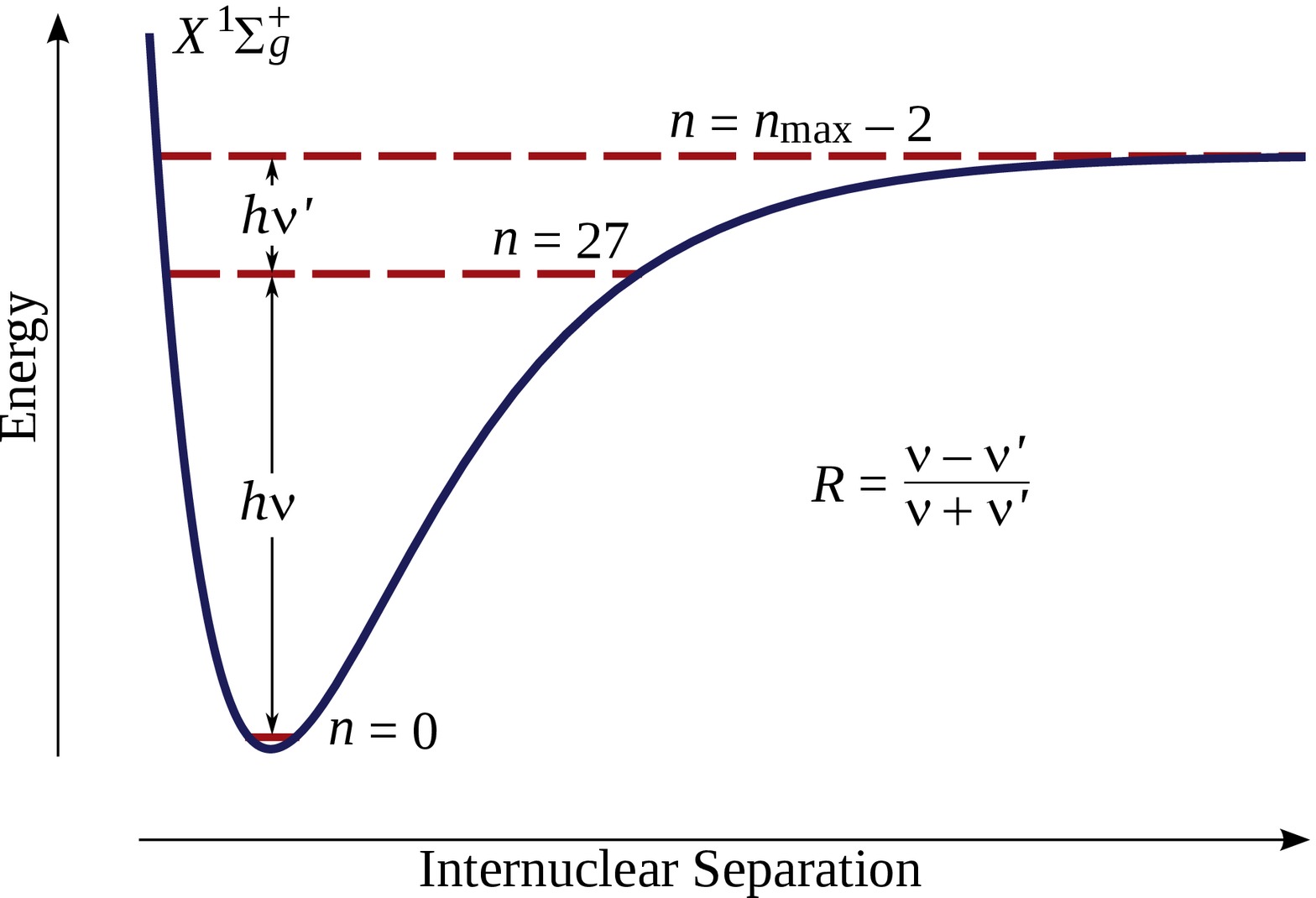}
\end{center}
\caption{(color online) Basic illustration of the experiment described by \zel~\cite{ZelKotYe08,KotZelYe09} to test the stability of the \ep{} using the vibrational spectrum of \sr{88}. The curve represents the \gx{} electronic potential, with horizontal dashed lines corresponding to select vibrational levels supported by this potential. Frequencies $\nu$ and $\nu^\prime$ are to be measured by an optical Ramsey scheme involving an intermediate excited electronic state (not shown). The dimensionless ratio $R$ is sensitive to variations in $\mu$ and is independent of any external reference (clock) frequency.}
\label{Fig:scheme}
\end{figure}

The Ye group at JILA (Boulder) aims to test the stability of $\mu$ to high precision using \sr{} molecules confined in an optical lattice \cite{Yeweb}. The experimental protocol, as outlined by \zel~\cite{ZelKotYe08,KotZelYe09}, calls for optical Raman spectroscopy between select vibrational levels of the \gx{} ground electronic potential. A variation in $\mu$ alters the vibrational spectrum, and experimental sensitivity to this change may be optimized with a prudent choice of levels to incorporate into the spectroscopic scheme. It was shown in Refs.~\cite{ZelKotYe08,DeMSaiSag08etal} that, with respect to variation in $\mu$, the lowest and highest vibrational levels experience minimal displacement relative to the potential, whereas levels in the intermediate part of the spectrum experience a much larger shift. \zel~have focused on transitions between the $n=27$ intermediate level ($n$ being the vibrational quantum number) and ``anchor'' levels at the bottom and top of the spectrum. Fig.~\ref{Fig:scheme} illustrates the basic objective of the experiment. A detected drift in the frequency ratio $R$ (see Figure) is to be interpreted as a drift in the \ep{}.

Here we investigate the effect of $\alpha$ variation on this promising experiment. The electronic potential depends on $\alpha$ through relativistic effects of electron motion.
A variation of $\alpha$ alters the potential and, consequently, the vibrational spectrum supported by it. Thus, a measured drift in $R$ may be due (or partially due) to $\alpha$ variation, threatening misinterpretation of the experimental results.

The experiment described by \zel{}~is similar in spirit to an experiment posed simultaneously by \dem~\cite{DeMSaiSag08etal} to test the stability of $\mu$ using diatomic molecules.
A key difference is that, whereas \zel{}~focus on a single electronic potential, \dem{}~suggest probing the splitting between vibrational levels supported by different electronic potentials. Recently, we analyzed the influence of $\alpha$ variation on the experiment of \dem~using a semi-classical (WKB) approach, specifically focusing on the system \cs{} \citep{BelBorFla11}. We found the experiment to be order-of-magnitude as sensitive to fractional variation in $\alpha$ as it is to fractional variation in $\mu$. Considering the anticipated precision of this experiment~\cite{DeMSaiSag08etal}, together with the current laboratory limit on $\alpha$ variation~\cite{RosHumSch08etal}, we concluded that $\alpha$ variation may not be negligible for the proposed experiment. This finding largely motivated our present work.

\section{Preliminary set-up}
Strictly speaking, only variations in dimensionless quantities have physical meaning. For a given vibrational level $n$, we will concern ourselves with the normalized energy $\mathcal{E}_n=E_n/D$, where $E_n$ is the vibrational energy relative to the bottom of the potential and $D$ is the potential depth. Clearly, $\mathcal{E}_n$ is limited to the range $0<\mathcal{E}_n<1$. Variations in $\mu$ and $\alpha$ induce a shift in $\mathcal{E}_n$,
\begin{eqnarray*}
\delta\mathcal{E}_n=\left(\partial_\mu \mathcal{E}_n\right)\frac{\delta\mu}{\mu}+\left(\partial_\alpha \mathcal{E}_n\right)\frac{\delta\alpha}{\alpha},
\end{eqnarray*}
where we employ the shorthand notation
\begin{eqnarray*}
\partial_\mu\equiv\frac{\partial}{\partial\ln\mu},
\qquad
\partial_\alpha\equiv\frac{\partial}{\partial\ln\alpha}.
\end{eqnarray*}
The quantities $\partial_\mu \mathcal{E}_n$ and $\partial_\alpha \mathcal{E}_n$ quantify the sensitivity of the vibrational level $n$ to fractional variations in the \ep{} and the fine structure constant, respectively.

We could, if desired, regard $\delta E_n=\delta\mathcal{E}_n\times D$ as an ``absolute'' energy shift, an association which amounts to arbitrarily assuming the potential depth to be fixed with respect to any variation.
Fixing any other energy reference---such as the atomic unit of energy, given by $e^4m_e/\hbar^2=\alpha^2 m_e c^2$, or the SI unit of energy, which itself references the hyperfine frequency of Cs as well as a platinum-iridium prototype mass held in Paris~\cite{TayTho08}---would be equally justified and would generally yield a different ``absolute'' energy shift.
Here we actively avoid the possibly slippery notion of absolute energy shift and quote results for $\partial_\mu \mathcal{E}_n$ and $\partial_\alpha \mathcal{E}_n$, as these are unambiguously defined.

\section{The Morse potential}

The Morse potential represents an idealized electronic potential for a diatomic molecule. It is given by
\begin{eqnarray*}
V(r)=D\left[1-e^{-a(r-r_0)}\right]^2,
\end{eqnarray*}
where $r$ is the internuclear separation, with $r_0$ being the equilibrium distance, and $a^{-1}$ is directly related to the width of the potential. The normalized vibrational energies for the Morse potential are given precisely by the formula
\begin{eqnarray}
\mathcal{E}_n=\epsilon\n-\frac{1}{4}\epsilon^2\n^2,
\label{Eq:MorseEnergies}
\end{eqnarray}
where $\epsilon\equiv\hbar a\sqrt{2/DM}$ and $M$ is the reduced nuclear mass.

\begin{figure}[t]
\begin{center}
\includegraphics*[scale=0.43]{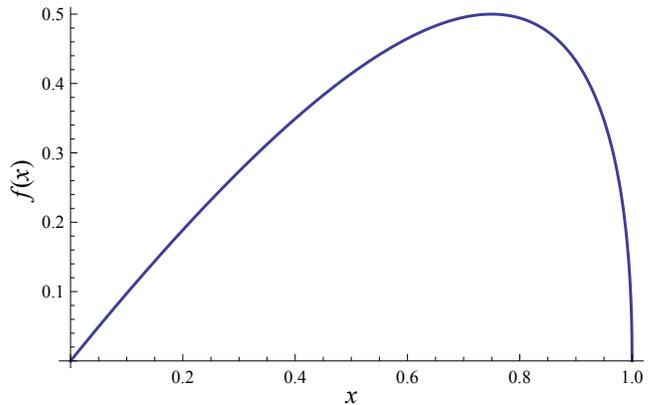}
\end{center}
\caption{The function $f(x)=2\left[x-1+\sqrt{1-x}\right]$. For the Morse potential, this function modulates the level sensitivities to both $\mu$ and $\alpha$ variation across the vibrational spectrum, with the argument $x$ taken as $\mathcal{E}_n\equiv E_n/D$. The most deeply ($\mathcal{E}_n\rightarrow0$) and loosely ($\mathcal{E}_n\rightarrow1$) bound levels are insensitive to variations, while intermediate levels have much larger sensitivities.}
\label{Fig:fz}
\end{figure}

From Eq.~(\ref{Eq:MorseEnergies}) we see that
a variation in $\mathcal{E}_n$ may be attributed solely to a variation in the parameter $\epsilon$. Specifically, we may write
\begin{eqnarray}
\delta\mathcal{E}_n=f(\mathcal{E}_n)\frac{\delta\epsilon}{\epsilon}=f(\mathcal{E}_n)\left[\left(\partial_\mu \ln\epsilon\right)\frac{\delta\mu}{\mu}+\left(\partial_\alpha \ln\epsilon\right)\frac{\delta\alpha}{\alpha}\right],
\nonumber\\
\label{Eq:Morsevar}
\end{eqnarray}
where $f(x)=2\left[x-1+\sqrt{1-x}\right]$. The function $f(x)$ is displayed in Fig.~\ref{Fig:fz}. This function modulates the sensitivity of the various levels of the vibrational spectrum to variations in $\mu$ and $\alpha$. Notably it approaches zero in the limits $x\rightarrow0$ and $x\rightarrow1$ and has a maximum at $x=3/4$. This translates to minimal sensitivities for the lowest and highest vibrational levels, with the largest sensitivities occurring for levels in the intermediate part of the spectrum.

We may go a step further and, based on physical reasoning, deduce a numerical value for the factor $(\partial_\mu \ln\epsilon)$ appearing in Eq.~(\ref{Eq:Morsevar}). This is accomplished most transparently by assuming atomic units, though we reiterate that $\epsilon$ itself is dimensionless. When expressed in atomic units, the molecular potential (and its depth, width, etc.)~is independent of the \ep{}, whereas the reduced mass has a value which is inversely proportional to $\mu$. From the definition of $\epsilon$, it follows that $(\partial_\mu \ln\epsilon)=1/2$.

In contrast to $(\partial_\mu \ln\epsilon)$, there is not a simple analytical result for the factor $(\partial_\alpha \ln\epsilon)$. Nevertheless, we may provide a physically reasonable estimate of $(\partial_\alpha \ln\epsilon)$ by realizing that, in atomic units, the electronic potential is independent of $\alpha$ in the nonrelativistic limit, having relativistic corrections which scale as $(\alpha Z)^2$, with $Z$ being the nuclear charge number. This suggests that $(\partial_\alpha \ln\epsilon)\sim(\alpha Z)^2$.
From this reasoning, we may suspect the vibrational spectrum of \sr{} to be nearly as sensitive to $\alpha$ variation as it is to $\mu$ variation.

The exercise of this section provides us with useful insight which is applicable to real physical systems. A true potential, of course, is not restricted to the form of a Morse potential. Nevertheless, the function $f(x)$ displayed in Fig.~\ref{Fig:fz} is expected to give a qualitatively accurate depiction of the sensitivities $\partial_\mu \mathcal{E}_n$ and $\partial_\alpha \mathcal{E}_n$ versus the normalized energy $\mathcal{E}_n$. In the vicinity of the equilibrium distance, the potential resembles that of a harmonic oscillator. The lower portion of the energy spectrum is then well-described by a single term, proportional to \n{}, in a Dunham-type expansion. Across this region, $\partial_\mu \mathcal{E}_n$ vs $\mathcal{E}_n$ and $\partial_\alpha \mathcal{E}_n$ vs $\mathcal{E}_n$ are essentially linear. Approaching the dissociation limit,
anharmonic effects become important and the remaining terms in the Dunham expansion, proportional to $\n^2$, $\n^3$, etc., then drive the sensitivities back to zero. The fact that the sensitivities approach zero in the limits $\mathcal{E}_n\rightarrow0$ and $\mathcal{E}_n\rightarrow1$ is a consequence of our choice for the zero of energy (bottom of the potential) and our choice to normalize energy to the dissociation energy.

\section{\abi{} calculations for level sensitivities}

We have calculated the \gx{} potential of \sr{}
using the relativistic computation chemistry program DIRAC10~\cite{DIRAC10}. In order to
reduce computational effort we employed the infinite order two-component
relativistic Hamiltonian obtained after the Barysz-Sadlej-Snijders (BSS)
transformation of the Dirac Hamiltonian in a finite basis set
\cite{IliJenKel05}. This approximation includes both scalar and spin-orbit
relativistic effects to infinite order and is one of the most computer
time efficient and accurate approximations to the four-component
Dirac-Coulomb Hamiltonian.  Electron correlation was taken into account
using closed-shell single-reference coupled-cluster theory including
single, double, and perturbative triple excitations [CCSD(T)]. The
Faegri dual family basis set \cite{Fae01} was used, augmented by diffuse
and high angular momentum exponents to obtain $21s18p12d6f2g$ Gaussian
orbitals. Virtual orbitals with energies above 45 a.u.~were omitted, and
the 56 outer core electrons were correlated.

We subsequently fed our CCSD(T) potential curve into a Matlab
routine to solve the Schr\"{o}dinger equation for the nuclear part of the
molecular wave function within the Born-Oppenheimer approach. A symmetric
three-point finite difference method was applied to obtain the nuclear
eigenfunctions together with their corresponding vibrational energies.
For the discretization of the internuclear distance a step size of
$7\times{}10^{-3}$~\AA{} was chosen.

We may gauge the accuracy of our \abi{} method through direct comparison with experimental results of Gerber \etal~\cite{GerMolSch84}. These authors have tabulated energies for the $n=0$ through $n=35$ portion of the \gx{} vibrational spectrum. Furthermore, they have determined the dissociation energy of this state
to be $D=1060(30)~\cm$. Our computed dissociation energy, $D=993~\cm$, is about $2\sigma$ lower than the experimental value. Comparing individual levels, we find that our computed vibrational energies differ from experimental values by no more than 2~\cm{} for levels spanning the lower half of the potential depth ($n=0$ through $n=15$). Above this, our computed energies steadily diverge from experiment values, with our values being increasingly smaller in comparison. For example, for $n=27$ our computed energy $E_n=789~\cm$ is 3\% lower than the experimental value $E_n=811~\cm$, whereas for $n=35$ our computed energy $E_n=889~\cm$ is 5\% lower than the experimental value $E_n=940~\cm$.
This divergence in the upper part of the spectrum is undoubtedly correlated to the fact that our dissociation energy is lower than the experimental dissociation energy.

%

\begin{figure}[t]
\begin{center}
\includegraphics*[scale=0.47]{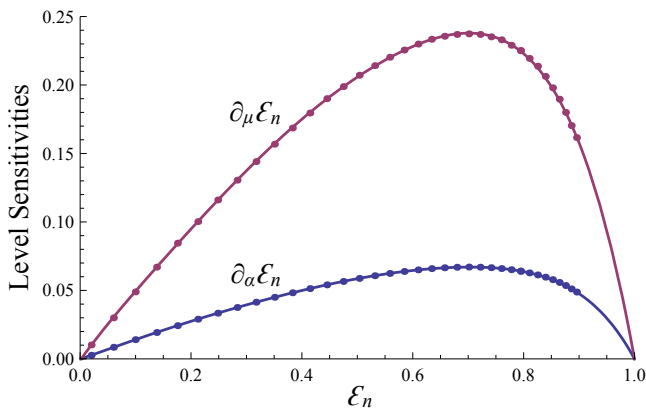}
\end{center}
\caption{(Color online) Level sensitivities $\partial_\mu \mathcal{E}_n$ and $\partial_\alpha \mathcal{E}_n$ vs $\mathcal{E}_n$ for the \gx{} state of \sr{}. The circles correspond to \abi{} results for individual levels $n=0$ through $n=35$, with the solid curves being fits to this data (the fits are extended through $\mathcal{E}_n=1$).}
\label{Fig:sensyandy}
\end{figure}

To see how the normalized energies change with respect to variations in $\mu$ and $\alpha$, we recompute the potential energy curve, as well as the vibrational spectrum supported by it, for various values of $\mu$ and $\alpha$ in the neighborhood of $\mu=1/1836$ and $\alpha=1/137$. The computational chemistry programs assume atomic units; numerical variations in $\mu$ and $\alpha$ are effected by modifying parameter values for the reduced mass within our Matlab routine and the speed of light within DIRAC10 ($M=44\,\mu^{-1}~\au$~for \sr{88} and $c=1/\alpha~\au$, where \au~denotes the respective atomic units of mass and velocity). We then obtain the sensitivities $\partial_\mu \mathcal{E}_n$ and $\partial_\alpha \mathcal{E}_n$ from numerical differentiation with respect to $\mu$ and $\alpha$. We emphasize that our method for obtaining these sensitivities treats variations in $\mu$ and $\alpha$ in a similar manner and on equal footing. Figure~\ref{Fig:sensyandy} displays our results for $\partial_\mu \mathcal{E}_n$ and $\partial_\alpha \mathcal{E}_n$ for the levels $n=0$ through $n=35$; these level sensitivities are plotted versus the normalized energy $\mathcal{E}_n$. We note a behavior which resembles that ``predicted'' by the Morse potential. Namely, both sensitivity curves approach the appropriate limits for $\mathcal{E}_n\rightarrow0$ and $\mathcal{E}_n\rightarrow1$, while simultaneously peaking at $\mathcal{E}_n\cong3/4$. Moreover, the curves are found to be essentially proportional, with the ratio $\partial_\alpha \mathcal{E}_n/\partial_\mu \mathcal{E}_n$ being $0.28\pm0.02$ across the entire range of data (and 0.28 at the common maximum of the two curves).

\section{Conclusion}
Here we have considered the influence of $\alpha$ variation on the experiment proposed by \zel~\cite{ZelKotYe08,KotZelYe09} to probe variation in the \ep{} using the vibrational spectrum of \sr{}. The relevant observable in this experiment is the frequency ratio $R$, illustrated in Fig.~\ref{Fig:scheme}.
With one ``anchor'' level taken at the bottom of the spectrum and another at the top, the frequency ratio is given approximately by
\begin{eqnarray*}
R\cong2\mathcal{E}_n-1,
\end{eqnarray*}
where $n$ labels the intermediate level (e.g., $n=27$ in Fig.~\ref{Fig:scheme}). Our \abi{} computations predict that the frequency ratio $R$ is only 3 to 4 times less sensitive to variation in $\alpha$ as it is to variation in $\mu$. Specifically, we find that variations in $\mu$ and $\alpha$ induce a variation in $R$ according to the relation
\begin{eqnarray}
\delta R=K\left(\frac{\delta\mu}{\mu}+0.28\frac{\delta\alpha}{\alpha}\right).
\label{Eq:deltaR}
\end{eqnarray}
We estimate the uncertainty in the factor 0.28 to be on the order of 10\%, based primarily on the discrepancy of our computed dissociation energy to the experimental dissociation energy. Equation (\ref{Eq:deltaR}) summarizes the principle result of this work. The factor $K$ here is given approximately by $K\cong2\times\partial_\mu\mathcal{E}_n$; non-zero shifts in the two anchor levels amount to small corrections which reduce $K$ from this value.

As suggested by Eq.~(\ref{Eq:deltaR}), a measured drift in the frequency ratio $R$ cannot, by itself, be used to distinguish between $\mu$ variation or $\alpha$ variation. To extract information about variations in the respective constants themselves requires further experimental input. Optical ion clocks have been used to test the stability of $\alpha$, with the ratio of clock frequencies being insensitive to $\mu$ variation. The current best limit on $\alpha$ variation allows for a drift on the fractional level of $4\times10^{-17}$/year \cite{RosHumSch08etal}. For the proposed experiment of \zel{}, this result may be used with Eq.~(\ref{Eq:deltaR}) to justify neglect of $\alpha$ variation, which has been implicitly assumed in previous works~\cite{ZelKotYe08,KotZelYe09}. On the other hand, for high experimental precision---namely, experimental precision capable of detecting a drift in $\mu$ at the fractional level of $1\times10^{-17}$/year---equation~(\ref{Eq:deltaR}) indicates that $\alpha$ variation should not be neglected. Such high precision is conceivable; in the related proposal of \dem~\cite{DeMSaiSag08etal}, referred to in the Introduction, the authors argued that their method could plausibly detect fractional variations in $\mu$ at $\lesssim10^{-17}$.

For such high experimental precision,
additional experimental input could perhaps be obtained by substituting \sr{} with another species, such as \yb{}, in the experiment. Yb has a similar valence structure as Sr and also has isotopes which lack nuclear spin ($^{168,170,172,174,176}$Yb). Moreover, as with Sr, high precision spectroscopy on optically trapped Yb has become a refined art \cite{LemLudBar09etal,PolBarLem08etal}.
From the $(\alpha Z)^2$ scaling of the relativistic corrections to the electronic potential, we may presume that an \yb{} experiment would be about equally sensitive to $\alpha$ variation as to $\mu$ variation, with an estimated sensitivity ratio $0.28\times(70/38)^2=0.95$. Using \sr{} and \yb{} results in conjunction, one could conceivably determine both $\mu$ and $\alpha$ variation to high precision with the proposed experiment of \zel


\section{Acknowledgements}
This work was supported by the Marsden Fund, administered by the Royal Society of New Zealand. VF further acknowledges support by the ARC.


%

\end{document}